\documentclass[12pt]{iopart}
\usepackage{setstack}
\usepackage{graphicx}
\usepackage{amssymb}
\begin{document}
\title{Jerk, snap, and the cosmological equation of state}
\author{Matt Visser}
\address{School of Mathematical and Computing Sciences, 
Victoria University of Wellington, PO Box 600, Wellington, New Zealand}
\ead{matt.visser@mcs.vuw.ac.nz}
\begin{abstract}
\def\d{{\mathrm{d}}}
  Taylor expanding the cosmological equation of state around the
  current epoch
  \[
  p = p_0 + \kappa_0\; (\rho-\rho_0) 
      + {1\over2} \left.{\d^2 p\over\d\rho^2}\right|_0  (\rho-\rho_0)^2 
      + O[(\rho-\rho_0)^3],
  \]
  is the simplest model one can consider that does not make any
  \emph{a priori} restrictions on the nature of the cosmological
  fluid.  Most popular cosmological models attempt to be
  ``predictive'', in the sense that once some \emph{a priori} equation
  of state is chosen the Friedmann equations are used to determine the
  evolution of the FRW scale factor $a(t)$. In contrast, a
  ``retrodictive'' approach might usefully take observational data
  concerning the scale factor, and use the Friedmann equations to
  \emph{infer} an observed cosmological equation of state.  In
  particular, the value and derivatives of the scale factor determined
  at the current epoch place constraints on the value and derivatives
  of the cosmological equation of state at the current epoch.
  Determining the first three Taylor coefficients of the equation of
  state at the current epoch requires a measurement of the
  deceleration, jerk, and snap --- the \emph{second}, \emph{third},
  and \emph{fourth} derivatives of the scale factor with respect to
  time.  Higher-order Taylor coefficients in the equation of state are
  related to higher-order time derivatives of the scale factor. Since
  the jerk and snap are rather difficult to measure, being related to
  the \emph{third} and \emph{fourth} terms in the Taylor series
  expansion of the Hubble law, it becomes clear why direct
  observational constraints on the cosmological equation of state are
  so relatively weak; and are likely to remain weak for the
  foreseeable future.

\vskip 0.50cm
\noindent
  Dated: 24 March 2004; \LaTeX-ed \today
\end{abstract}
\pacs{gr-qc/0309109}
\maketitle
\newtheorem{theorem}{Theorem}
\newtheorem{corollary}{Corollary}
\newtheorem{lemma}{Lemma}
\def\d{{\mathrm{d}}}
\def\implies{\Rightarrow}

\def\eg{{\it e.g.}}
\def\etc{{\it etc}}
\def\sign{{\hbox{sign}}}
\def\eof{\Box}
\newenvironment{warning}{{\noindent\bf Warning: }}{\hfill $\eof$\break}
\section{Introduction}

This article develops a ``phenomenological'' approach to the equation
of state [EOS] of the cosmological fluid, and investigates what would
have to be done in order to observationally determine the EOS.  Even
at the linearized level, where
\begin{equation}
p = p_0 + \kappa_0\; (\rho-\rho_0) + O[(\rho-\rho_0)^2],
\end{equation}
the first nontrivial coefficient in the EOS will be seen to be related
to the cosmological jerk --- the \emph{third} derivative of the scale
factor with respect to time, and thence to the \emph{third}-order term
in the Taylor series expansion of the Hubble law. This is the
fundamental reason why observational determinations of the EOS are
relatively poor, and why it is possible to choose so many wildly
differing
\emph{a priori} models for the EOS that nevertheless give good
agreement with the coarse features of the present epoch.

More generally, if we describe the cosmological equation of state in
terms of a Taylor series expansion around the current epoch
\begin{equation}
  p = p_0 + 
  \sum_{n=1}^{N-1}  {1\over n!} 
  \left.{\d^n p\over\d\rho^n}\right|_0  (\rho-\rho_0)^n 
  + O[(\rho-\rho_0)^N],
\end{equation}
then the $n$-th order Taylor coefficient 
\begin{equation}
\left.{\d^n p\over\d\rho^n}\right|_0
\end{equation}
will be shown to depend on the ($n$+2)-th time derivative of the scale
factor
\begin{equation}
\left.{\d^{n+2} a(t)\over\d t^{n+2}}\right|_0,
\end{equation}
and thence to the ($n$+2)-th term, the $O(z^{n+2})$ term, in the
Taylor expansion of the Hubble law.

In most attempts at cosmological model building one takes a FRW
cosmology
\begin{equation}
\d s^2 = - c^2 \;\d t^2 
+ a(t)^2 \left[ {\d r^2\over 1- k \,r^2} 
+ r^2\left(\d\theta^2+\sin^2\theta\;\d\phi^2\right) \right];
\end{equation}
plus the conservation of stress-energy 
\begin{equation}
\dot \rho \; a^3 + 3[\rho+p] \; a^2 \dot a = 0;
\end{equation}
and chooses some \emph{a priori} equation of state
\begin{equation}
\rho = \rho(p); \qquad \hbox{or} \qquad p = p(\rho);
\end{equation}
to derive $\rho(a)$, and equivalently $p(a)$. The Einstein equations
then reduce to the single Friedmann equation, which can be written in
the form
\begin{equation}
\dot a = \sqrt{ {8\pi G_N \; \rho(a) \; a^2 \over 3} - k },
\end{equation}
and used to determine $a(t)$. (See, for example, any standard text
such as~\cite{Weinberg,MTW,Peebles}.)

In contrast, let us assume we have a FRW universe with good
observational data on $a(t)$ --- in Weinberg's terminology we have a
good ``cosmography''~\cite{Weinberg}.  In this situation we can use
the Einstein equations in reverse to calculate the energy density
$\rho(t)$ and pressure $p(t)$ via
\begin{equation}
8\pi G_N \; \rho(t) =
3  c^2 \left[{\dot a^2\over a^2} + {k c^2\over a^2} \right];
\end{equation}
\begin{equation}
8\pi G_N \; p(t) =
- c^2 \left[ 
{\dot a^2\over a^2} + {k c^2\over a^2} + 2 {\ddot a\over a}
\right].
\end{equation}
Under mild conditions on the existence and nonzero value of
appropriate derivatives we can appeal to the inverse function theorem
to assert the existence of a $t(\rho)$ or $t(p)$ and hence, in
principle, deduce an observational equation of state
\begin{equation}
\rho(p) = \rho(t=t(p)); \qquad p(\rho) = p(t=t(\rho)).
\end{equation}
In view of the many controversies currently surrounding the
cosmological equation of state, and the large number of speculative
models presently being considered, such an observationally driven
reconstruction is of interest in its own right.

Now in observational cosmology we do not have direct access to $a(t)$
over the entire history of the universe --- we do however have access
[however imprecise] to the current value of the scale factor and its
derivatives, as encoded in the Hubble parameter, deceleration
parameter, \etc. This more limited information can still be used to
extract useful information about the cosmological equation of state,
in particular it yields information about the present value of the
$w$-parameter and the slope parameter $\kappa_0$ defined as
\begin{equation}
\label{E:kappa}
w_0 = \left.{p\over\rho}\right|_0; 
\qquad
\kappa_0 = \left.{\d p\over\d\rho}\right|_0.
\end{equation}
The value of the $w$-parameter in particular has recently become the
center of considerable interest, driven by speculation that $w_0<-1$
is compatible with present observations.  Such a value of $w_0$ would
correspond to present-day classical and cosmologically significant
violations of the null energy condition. The associated ``phantom
matter'' (\emph{almost} identical to the notion of ``exotic matter''
in the sense of Morris and Thorne~\cite{Morris-Thorne}) leads to a
cosmological energy density that is future increasing rather than
future decreasing.  (See, for example,~\cite{Bounce,Tolman}).  If
$w(t)$ subsequently remains strictly less than $-1$, this will lead to
a ``big rip''~\cite{big-rip} --- the catastrophic infinite expansion
of the universe in finite elapsed time.

Unfortunately it is very difficult to measure $w_0$ and $\kappa_0$
with any accuracy --- I will make this point explicit by relating the
measurement of $w_0$ to the deceleration parameter, and the
measurement of $\kappa_0$ to the ``jerk'' of the cosmological scale
factor --- the third derivative with respect to time. 

For related comments see references~\cite{Chiba,S1,S2,S3,P1,P2}. The
``cubic'' term of Chiba and Nakamura~\cite{Chiba} is identical to the
jerk, as is the ``statefinder'' variable called $r$ by Sahni~\emph{et
  al.}~\cite{S1,S2,S3}. The other ``statefinder'' variable (called
$s$, not to be confused with the snap) is a particular linear
combination of the jerk and deceleration parameters.  Padmanabhan and
Choudhury~\cite{P1} have also emphasised the need for constructing
models for the cosmological fluid that are unprejudiced by \emph{a
  priori} theoretical assumptions. A good recent survey of the status
of the cosmological fluid is~\cite{P2}.

\section{Hubble, deceleration, jerk, and snap  parameters}

It is standard terminology in mechanics that the first four time
derivatives of position are referred to as velocity, acceleration,
jerk and snap.~\footnote{Jerk [the third time derivative] is also
  sometimes referred to as jolt.  Less common alternative
  terminologies are pulse, impulse, bounce, surge, shock, and
  super-acceleration. Snap [the fourth time derivative] is also
  sometimes called jounce. The fifth and sixth time derivatives are
  sometimes somewhat facetiously referred to as crackle and pop.}  In
a cosmological setting this makes it appropriate to define Hubble,
deceleration, jerk, and snap parameters as
\begin{equation}
H(t) = + {1\over a} \; {\d a\over\d t};
\end{equation}
\begin{equation}
q(t) = - {1\over a} \; {\d^2 a\over \d t^2}  \;
\left[ {1\over a} \; {\d a \over  \d t}\right]^{-2};
\end{equation}
\begin{equation}
j(t) = + {1\over a} \; {\d^3 a \over \d t^3}  
\; \left[ {1\over a} \; {\d a \over  \d t}\right]^{-3};
\end{equation}
\begin{equation}
s(t) = + {1\over a} \; {\d^4 a \over \d t^4}  
\; \left[ {1\over a} \; {\d a \over  \d t}\right]^{-4}.
\end{equation}
The deceleration, jerk, and snap parameters are dimensionless, and we
can write
\begin{eqnarray}
\fl
a(t)= a_0 \;
\Bigg\{ 1 + H_0 \; (t-t_0) - {1\over2} \; q_0 \; H_0^2 \;(t-t_0)^2 
+{1\over3!}\;  j_0\; H_0^3 \;(t-t_0)^3 
\nonumber
\\
\qquad
+{1\over4!}\;  s_0\; H_0^4 \;(t-t_0)^4
+ O([t-t_0]^5) \Bigg\}.
\end{eqnarray}
In particular, at arbitrary time $t$
\begin{equation}
w(t) = {p\over\rho} 
= 
- {H^2(1-2q) + k c^2/a^2 \over 3(H^2 + k c^2/a^2)}
= 
- {(1-2q) + k c^2/(H^2a^2) \over 3[1 + k c^2/(H^2a^2)]}.
\end{equation}
While observation is currently not good enough to distinguish between
the three cases $k=-1/0/+1$ with any degree of certainty, there is
nevertheless widespread agreement that at the present epoch $H_0 a_0/c
\gg 1$ (equivalent to $|\Omega_0-1| \ll 1$).

\begin{warning}
  From a theoretical perspective, $H_0 a_0/c \gg 1$ is a generic
  prediction of inflationary cosmology --- this is \emph{not the same}
  as saying that cosmological inflation predicts $k=0$. What generic
  cosmological inflation predicts is the weaker statement that for all
  practical purposes the \emph{present day} universe is
  indistinguishable from a $k=0$ spatially flat universe.  If our
  universe happens to be a topologically trivial $k=0$ FRW cosmology,
  then \emph{we will never be able to prove it}.  Simply as a matter
  of formal logic, all we will ever be able to do is to place
  increasingly stringent lower bounds on $H_0\, a_0$, but this will
  never rigorously permit us to conclude that $k=0$. The fundamental
  reason for this often overlooked but trivial observation is that a
  topologically trivial $k=0$ FRW universe can be mimicked to
  arbitrary accuracy by a $k=\pm1$ FRW universe provided the scale
  factor is big enough.~\footnote{If the universe has nontrivial
  spatial topology there is a possibility of using the
  compactification scale, which might be (but does not have to be)
  much smaller than the scale factor, to indirectly distinguish
  between $k=-1/0/+1$.}  In contrast if the true state of affairs is
  $k=\pm1$, then with good enough data on $H_0 a_0$ we will in
  principle be able to determine upper bounds which (at some
  appropriate level of statistical uncertainty) demonstrate that
  $k\neq0$.  Also note that even in inflationary cosmologies it is not
  true that $H(t) a(t)/c \gg 1$ at all times, and in particular this
  inequality \emph{may} be violated (and often is violated) in the
  pre-inflationary epoch.
\end{warning}
Now the $w$-parameter in cosmology is related to the Morris--Thorne
exoticity parameter~\cite{Morris-Thorne} which was introduced by them
to characterize the presence of ``exotic matter'', matter violating
the null energy condition [NEC]:
\begin{equation}
\xi = {\rho+p\over |\rho|} = \sign(\rho) \; [1+w]
= {2\over3}\; \sign(\rho) \; 
{1+q + kc^2/(H^2 a^2)\over 1 + kc^2/(H^2 a^2)}
\end{equation}
Thus if $w<-1$ and $\rho>0$ we have $\xi<0$ and the NEC is violated.
In contrast, if $w<-1$ but $\rho< 0$ we have $\xi>0$, the NEC is
satisfied but the weak energy condition [WEC] is violated. That is,
``phantom matter'' (matter with $w<1$) is not quite the same as
``exotic matter'' (for which $\xi<0$), but the two are intimately
related.

Accepting the approximation that $H_0 a_0/c \gg 1$ we have
\begin{equation}
\rho_0 \approx {3\over8\pi G_N}\; H_0^2 > 0; \quad
w_0 \approx - {(1-2q_0)\over 3}; 
\qquad \hbox{and} \quad
\xi_0 \approx {2\over3}\;(1+q_0);
\end{equation}
so that in this situation the $w_0$-parameter and exoticity parameter
$\xi_0$ are intimately related to the deceleration parameter $q_0$. In
particular if $w_0<-1$, so that the universe is at the current epoch
dominated by ``phantom matter'', we also (because in this
approximation $\rho_0$ is guaranteed to be positive) have $\xi_0<0$ so
that at the current epoch this phantom matter is also ``exotic
matter''.  Exotic matter is powerful stuff: Apart from possibly
destroying the universe in a future ``big rip''
singularity~\cite{big-rip}, if the exotic matter clumps to any extent
there is real risk of even more seriously bizarre behaviour ---
everything from violations of the positive mass condition (that is,
objects with negative asymptotic mass), through traversable wormholes,
to time
warps~\cite{Morris-Thorne,Natural,Lorentzian,Twilight,Chronology}.

\section{Taylor series equation of state}

Linearize the cosmological EOS around the present epoch as
\begin{equation}
p = p_0 + \kappa_0\; (\rho-\rho_0) + O[(\rho-\rho_0)^2].
\end{equation}
To calculate $\kappa_0$ we use
\begin{equation}
 \kappa_0 =
{\left.\d p/\d t\right|_0 \over \left.\d \rho /\d t\right|_0},
\end{equation}
where numerator and denominator can be obtained by differentiating the
Friedmann equations for $\rho(t)$ and $p(t)$.  It is easy to see that
at all times, simply from the definition of deceleration and jerk
parameters, we have
\begin{equation}
8\pi G_N \; {\d\rho\over\d t} = -6 c^2 H \left[(1+q)H^2 + {kc^2\over a^2} \right],
\end{equation}
\begin{equation}
8\pi G_N \; {\d p\over\d t} = 2 c^2 H \left[(1-j)H^2 + {kc^2\over a^2} \right], 
\end{equation}
leading to
\begin{equation}
\kappa_0 = - {1\over 3} \;  \left[
{
1-j_0 + kc^2/(H_0^2a_0^2)
\over
1+q_0+ kc^2/(H_0^2 a_0^2)
} \right],
\end{equation}
which approximates (using  $H_0 a_0/c \gg 1$)  to
\begin{equation}
\kappa_0 = - {1\over 3} \;  \left[
{
1-j_0
\over
1+q_0
} \right]. 
\end{equation}
The key observation here is that to obtain the linearized equation of
state you need significantly more information than the deceleration
parameter $q_0$; you also need to measure the jerk parameter $j_0$.
If the only observations you have are measurements of the deceleration
parameter then you can of course determine $w_0=p_0/\rho_0$, \emph{but
  this is not an equation of state for the cosmological fluid}.
Determining $w_0$ merely provides information about the present-day
value of $p/\rho$ but makes no prediction as to what this ratio will
do in the future --- not even in the near future. (This point is also
forcefully made in~\cite{P1}.)  For this reason there have been
several attempts to observationally determine $w(z)$, the value of $w$
as a function of redshift.  See for example~\cite{P1}
and~\cite{S1,S2,S3}.  Since $z$ is a function of lookback time $D/c$,
this is ultimately equivalent to determining $w(t) = p(t)/\rho(t)$,
and implicitly equivalent to reconstructing a phenomenological
equation of state $p(\rho)$.  I prefer to phrase the discussion
directly in terms of the EOS as that will make it clear what
parameters have to be physically measured.  In terms of the history of
the scale factor $a(t)$, it is only when one goes to third order by
including the jerk parameter $j_0$ that one obtains even a linearized
equation of state.

Going one step higher in the expansion, by using the chain rule and
the implicit function theorem it is easy to see that
\begin{equation}
{\d^2 p\over\d\rho^2}  = 
{\ddot p- \kappa \; \ddot \rho\over (\dot\rho)^2}
\end{equation}
More generally ${\d^n p/\d\rho^n}$ contains a term linear in ${\d^n
p/\d t^n}$. Using the Friedmann equations then implies that ${\d^n
p/\d\rho^n}$ contains a term linear in ${\d^{n+2} a/\d t^{n+2}}$.
Specifically for the first nonlinear term it is relatively
straightforward take explicit time derivatives and so to verify that
\begin{eqnarray}
\fl
\left.{\d^2 p\over\d\rho^2}\right|_0  = 
-{(1+kc^2/[H_0^2 a_0^2]) \over 6 \rho_0 (1+q_0+kc^2/[H_0^2 a_0^2])^3}
\Big\{s_0(1+q_0)+j_0(1+j_0+4q_0+q_0^2)+q_0(1+2q_0) 
\nonumber
\\
+ 
(s_0+j_0 +q_0 + q_0 j_0){k c^2\over H_0^2 a_0^2}\Big\}.
\end{eqnarray}
In the approximation $H_0 a_0/c \gg 1$ this reduces to
\begin{equation}
\left.{\d^2 p\over\d\rho^2}\right|_0  =  
-{s_0(1+q_0)+j_0(1+j_0+4q_0+q_0^2)+q_0(1+2q_0)
\over 6 \rho_0 (1+q_0)^3}.
\end{equation}
As expected, this second derivative depends linearly on the snap.
Higher order coefficients can certainly be computed but are
increasingly complicated and less transparent in their physical
interpretation. (Calculations are impractical without the use of some
symbolic manipulation package such as Maple.) To now make the
connection between the the Taylor coefficients of the cosmological EOS
and the various parameters appearing in the Hubble law we will need to
likewise perform a similar Taylor expansion of the Hubble law.

\section{Hubble law to fourth order in redshift}

Note that this entire section is independent of the use of the
Friedmann equations and depends only on the use of a FRW geometry.

The physical distance travelled by a photon that is emitted at time
$t_*$ and absorbed at the current epoch $t_0$ is
\begin{equation}
D = c \int \d t = c\;(t_0 - t_*).
\end{equation}
In terms of this physical distance the Hubble law is \emph{exact}
\begin{equation}
1 + z = {a(t_0)\over a(t_*)} = {a(t_0)\over a(t_0 - D/c)},
\end{equation}
but impractical. A more useful result is obtained by performing a
fourth-order Taylor series expansion,
\begin{eqnarray}
\fl
{a(t_0)\over a(t_0 - D/c)} =
1 + {H_0 D\over c} +{2+q_0\over2} \; {H_0^2 D^2\over c^2} 
+ 
{6(1+q_0)+j_0\over6} \; {H_0^3 D^3\over c^3} 
\nonumber
\\
+{24-s_0+8j_0+36q_0+6q_0^2\over24} \;
 {H_0^4 D^4\over c^4} +
 O\left[\left(H_0 D\over c\right)^5\right],
\end{eqnarray}
followed by reversion of the resulting series $z(D)\to D(z)$ to
obtain:
\begin{eqnarray}
\fl
D = {c\; z\over H_0}
\Bigg\{ 1 
- 
\left[1+{q_0\over2}\right] {z} 
+
\left[ 1 + q_0 + {q_0^2\over2} - {j_0\over6}   \right] z^2 
\\
-
\left[ 1 +{3\over2}q_0(1+q_0)+{5\over8}q_0^3-{1\over2}j_0 
- {5\over12} q_0 j_0 -{s_0\over24} \right] z^3
+  O(z^4) \Bigg\}.
\nonumber
\label{E:physical}
\end{eqnarray}
This simple calculation is enough to demonstrate that the jerk shows
up at third order in the Hubble law, and the snap at fourth
order. Generally, the $O(z^n)$ term in this version of the Hubble law
will depend on the $n$-th time derivative of the scale factor.  (Also
note that one of the virtues of this version of the Hubble law is that
it is completely independent of $k$, the sign of space curvature.)

Unfortunately physical distance $D$ is typically not the variable in
terms of which the Hubble law is observationally presented. That role
is more typically played by the ``luminosity distance'', $d_L$. For
instance, Weinberg defines~\cite{Weinberg}
\begin{equation}
\hbox{(energy flux)} = {L\over4\pi \; d_L^2}.
\end{equation}
Let the photon be emitted at $r$-coordinate $r=0$ at time $t_*$, and
absorbed at $r$-coordinate $r=r_0$ at time $t_0$. Then it is a purely
geometrical result that
\begin{equation}
d_L = a(t_0)^2 \; {r_0\over a(t_*)} = {a_0\over a(t_0-D/c)} \; (a_0\,r_0).
\end{equation}
Thus to calculate $d_L(D)$ we need $r_0(D)$.  Recall that for a null
geodesic in a FRW universe
\begin{equation}
\int_{t_*}^{t_0} {c \;\d t \over a(t)} = 
\int_0^{r_0} {\d r\over \sqrt{1-k r^2}} = f(r_0).
\end{equation}
But
\begin{equation}
f(r_0) = \left\{ 
\begin{array}{ll}
\sin^{-1} r_0 & k=+1;\\
r_0 & k=0;\\
\sinh^{-1} r_0 \;\;& k=-1;
\end{array}
\right.
\end{equation}
therefore
\begin{equation}
r_0(D) = f^{-1} \left( \int_{t_*(D)}^{t_0} {c\;\d t \over a(t)} \right) =  
f^{-1} \left( \int_{t_0-D/c}^{t_0} {c\;\d t \over a(t)} \right).
\end{equation}
To be explicit
\begin{equation}
r_0(D) = \left\{
\begin{array}{cc}
\sin\left( \displaystyle\int_{t_0-D/c}^{t_0} {c\;\d t \over a(t)} \right)       & k=+1;\\
\\
\displaystyle\int_{t_0-D/c}^{t_0} {c\;\d t \over a(t)}                          & k=0;\\
\\
\sinh\left( \displaystyle\int_{t_0-D/c}^{t_0} {c\;\d t \over a(t)} \right) \;\; & k=-1.
\end{array}
\right.
\end{equation}
We can now Taylor series expand for ``short'' distances. First note that
\begin{equation}
\fl
r_0(D) = 
\left[\int_{t_0-D/c}^{t_0} {c\;\d t \over a(t)}\right]
- {k\over 3!} 
\left[ \int_{t_0-D/c}^{t_0} {c\;\d t \over a(t)} \right]^3 
+ O\left( \left[ \int_{t_0-D/c}^{t_0} {c\;\d t \over a(t)} \right]^5 \right),
\end{equation}
and now expand the integral to third order. (We can check, \emph{a
  posteriori}, that this retains sufficient accuracy in the $d_L
\leftrightarrow D$ conversion for determining the Hubble law to fourth
order.) Then
\begin{eqnarray}
\fl
\int_{t_0-D/c}^{t_0} {c\;\d t \over a(t)}
=\int_{t_0-D/c}^{t_0}  {c\;\d t \over a_0} \;
\Bigg\{1+ H_0(t_0-t) + \left[{2+q_0\over2} H_0^2 \right](t_0-t)^2
\nonumber\\
+
\left[{6(1+q_0)+j_0\over6} H_0^3\right](t_0-t)^3
+
O[(t_0-t)^4] \Bigg\}
\\
\nonumber
\\
\lo{=}
 {c\over a_0} \Bigg\{ 
D/c 
+{1\over2} H_0 (D/c)^2 
+\left[ {2+q_0\over6} \; H_0^2  \right](D/c)^3 
\nonumber\\
+\left[{6(1+q_0)+j_0\over24} \; H_0^3\right](D/c)^4
+ O\left[ H_0^4 \; (D/c)^5\right]\Bigg\}
\\
\nonumber
\\
\lo{=}
 {D\over a_0} \Bigg\{ 1 +{1\over2} {H_0 D\over c} 
+\left[ {2+q_0\over6} \right]
\left({H_0 D\over c}\right)^2
+\left[{6(1+q_0)+j_0\over24}\right]
\left({H_0 D\over c}\right)^3
\nonumber\\ 
+
O\left[\left(H_0 D\over c\right)^4\right]\Bigg\}.
\end{eqnarray}
The conversion from physical distance travelled to $r$ coordinate
traversed is given by
\begin{eqnarray}
\fl
r_0(D) = {D\over a_0} 
\Bigg\{ 1 +{1\over2} {H_0 D\over c} + 
 {1\over6} \left[ 2+q_0 - {kc^2\over H_0^2 a_0^2} \right] \; 
\left(H_0 D\over c\right)^2 
\nonumber
\\
+
{1\over24} \left[6(1+q_0)+j_0  - 6 {kc^2\over H_0^2 a_0^2}\right] 
\left(H_0 D\over c\right)^3
+
O\left[\left(H_0 D\over c\right)^4\right]\Bigg\}.
\end{eqnarray}
Combining these formulae we find that the luminosity distance as a
function of $D$, the physical distance travelled is:
\begin{eqnarray}
\fl
d_L(D) = {D}  \Bigg\{ 1  +{3\over2} \left({H_0 D\over c}\right) 
+
{1\over6}\left[11+4q_0-{kc^2\over H_0^2a_0^2}\right]
\left(H_0 D\over c\right)^2
\nonumber
\\
+
{1\over24}\left[50+40q_0+5j_0-10{k c^2\over H_0^2 a_0^2}    \right]
\left(H_0 D\over c\right)^3
+ O\left[\left(H_0 D\over c\right)^4\right]\Bigg\}.
\end{eqnarray}
Now using the series expansion for for $D(z)$ we finally derive the
luminosity-distance version of the Hubble law:
\begin{eqnarray}
\fl
d_L(z) =  {c\; z\over H_0}
\Bigg\{ 1 + {1\over2}\left[1-q_0\right] {z} 
-{1\over6}\left[1-q_0-3q_0^2+j_0+ {kc^2\over H_0^2a_0^2}\right] z^2
\nonumber
\\
\lo{+}
{1\over24}\left[
2-2q_0-15q_0^2-15q_0^3+5j_0+10q_0j_0+s_0 + 
{2 k c^2(1+3q_0)\over H_0^2 a_0^2}
\right] z^3
\nonumber
\\
+ O(z^4) \Bigg\}.
\end{eqnarray}
The first two terms above are Weinberg's version of the Hubble law.
His equation (14.6.8). The third term is equivalent to that obtained
by Chiba and Nakamura~\cite{Chiba}. The fourth order term appears to
be new, and (as expected) depends linearly on the snap. From the
derivation above it is now clear that the $O(z^n)$ term in this
luminosity distance version of the Hubble law will also depend on the
$n$-th time derivative of the scale factor.

It is important to realise that this Hubble law, and indeed the entire
discussion of this section, is completely model-independent --- it
assumes only that the geometry of the universe is well approximated by
a FRW cosmology but does not invoke the Einstein field equations
[Friedmann equation] or any particular matter model.  Note that in
comparison to the $D(z)$ Hubble law, this $d_L(z)$ Hubble law first
differs in the coefficient of the $O(z^2)$ term --- you will still get
the same Hubble parameter, but if you are not sure which definition of
``distance'' you are using you may mis-estimate the higher-order
coefficients (deceleration, jerk, and snap).  The jerk $j_0$ first
shows up in the Hubble law at \emph{third} order (order $z^3$); but
this was one of the parameters we needed to make the
\emph{lowest-order} estimate for the slope of the EOS. 

\begin{warning}
  Not all authors use the same definition of the luminosity distance.
  In particular D'Inverno uses a definition that differs from
  Weinberg's by an extra factor of $(1+z)^2$~\cite{D'Inverno}.
  Weinberg's definition as presented above appears to the most
  standard, but if necessary the conversion is straightforward.
\end{warning}

\section{A specific ``a priori'' model: Incoherent mixture of $w$-matter}

Though the philosophy so far has been to avoid committing ourselves to
any particular matter model, it is useful for comparison purposes to
see how these ideas impact on the most popular models.  A particularly
common \emph{a priori} model for the cosmological fluid is an
incoherent mixture of various forms of $w$-matter with each component
satisfying a zero-offset equation of state:
\begin{equation}
p_i = w_i \; \rho_i.
\end{equation}
Integrating the conservation equation independently for each component
of the mixture yields
\begin{equation}
\fl
p_i = p_{0i} \; (a/a_0)^{-3(1+w_i)} 
=  
\rho_{0i} \; w_i \; (a/a_0)^{-3(1+w_i)}
= 
\rho_c \; \Omega_{0i}  \; w_i \; (a/a_0)^{-3(1+w_i)}.
\end{equation}
This model is sufficiently general to contain dust, radiation,
cosmological constant, and standard forms of quintessence.  Then
\begin{equation}
w_0 = {\sum_i \Omega_{0i}  \; w_i \over \sum_i \Omega_{0i} } = \overline{w},
\end{equation}
is simply the weighted average value of $w$. (A sum of $w_i$ over all
matter components $i$, weighted by their present-day contribution to
the $\Omega$ parameter.)  Similarly
\begin{equation}
\kappa_0 =
{\left.\d p/\d a\right|_0 \over \left.\d \rho /\d a\right|_0}
=
{\sum_i \Omega_{0i}  \; w_i(1+w_i) 
\over \sum_i \Omega_{0i} \; (1+w_i) }
=
{\overline{w^2} + \overline{w} \over 1 +\overline{w}}
=
{\sigma^2_w  \over 1 +\overline{w}} + \overline{w}.
\end{equation}
That is, $w_0$ (and hence $q_0$) provides information about the
weighted average value of the $w_i$, while $\kappa_0$ (and hence
$j_0$) provides information about how much spread there is in the
various $w_i$. Uncertainties in the jerk parameter $j_0$ (which in a
generic model manifest themselves as uncertainties in the slope
parameter $\kappa_0$) in this specific model show up as difficulty in
determining the weighting parameters $\Omega_{0i}$. (For similar
comments, see~\cite{P1}.) Note that the positivity of $\sigma_w^2$
implies a constraint
\begin{equation}
(1 +\overline{w}) \; (\kappa_0 -\overline{w}) > 0.
\end{equation}
If this inequality observationally fails, it means that the
cosmological fluid cannot be described by any possible linear
combination of $w$-matter. In the approximation $H_0 a_0/c \gg 1$ this reduces to
\begin{equation}
j_0 > q_0(1-q_0).
\end{equation}

At the next highest level in the Taylor expansion, again making use of
the implicit function theorem, we have
\begin{equation}
{\d^2 p\over\d\rho^2}  = 
{p''- \kappa \; \rho''\over (\rho')^2}.
\end{equation}
After a brief computation this leads to
\begin{equation}
\rho_0 \; \left.{\d^2 p\over\d\rho^2}\right|_0  = 
{\overline{w^3} (1 +\overline{w}) +
\overline{w^2} (1-\overline{w}-\overline{w^2}) -
\overline{w}^2
\over (1 +\overline{w})^3}.
\end{equation}
In terms of standard deviation and skewness, 
\begin{equation}
{\cal S}_w= \overline{(w-\overline{w})^3} = 
\overline{w^3} - 3 \overline{w^2}\;\overline{w} +2 \overline{w}^3 =
\overline{w^3} - 3 \sigma^2_w - \overline{w}^3,
\end{equation}
 this yields
\begin{equation}
\rho_0 \; \left.{\d^2 p\over\d\rho^2}\right|_0  = 
{ {\cal S}_w  (1 +\overline{w}) - \sigma_w^4 
- 2\sigma_w^2 \;\overline{w}^2 - 2\sigma_w^2 
- 4 \sigma^2\;\overline{w} - 2 \overline{w}^3 
- 2 \overline{w}^4
\over (1 +\overline{w})^3}.
\end{equation}
Thus $\left.{\d^2 p/\d\rho^2}\right|_0$ is related to skewness in the
distribution of the $w_i$.  The general message to be extracted here
is that the $n$-th Taylor coefficient in the EOS depends linearly on
the ($n$+1)-th $\Omega$-weighted moment of the $w_i$.

\section{The observational situation}

As of March 2004 the observational situation is best summarized by the
dataset described in the recent article by Riess \emph{et
  al.}~\cite{Riess}. (See also the brief analysis in Caldwell and
Kamionkowski~\cite{Caldwell}). The Riess \emph{et al.} analysis
presupposes a $k=0$ spatially flat universe (effectively, they adopt
the inflationary paradigm and use the approximation $c/(H_0 a_0) \ll
1$). They report that the jerk $j_0$ is positive at the 92\%
confidence level based on their ``gold'' dataset, and is positive at
the 95\% confidence level based on their ``gold+silver'' dataset. No
explicit upper bounds are given for the jerk, nor are any constraints
placed on the snap $s_0$.  The allowed region (for their preferred
parameterization of the data in terms of $[\d q/\d z]_0$) is presented
in their Figure 5, and can usefully be summarized as follows: the
allowed region is a narrow ellipse approximately centred along the
line
\begin{equation}
{\d q\over\d z} = -1 -{7\over2}\; q_0
\end{equation}
and bounded by the boxes
\begin{equation}
\fl
q_0 \in (-1.3,-0.2); \qquad 
\left.{\d q\over\d z}\right|_0 \in (-0.2,+3.8) \qquad
\hbox{(99\% confidence, ``gold'')},
\end{equation}
and
\begin{equation}
\fl
q_0 \in (-1.4,-0.3); \quad 
\left.{\d q\over\d z}\right|_0 \in (+0.0,+3.9) \quad
\hbox{(99\% confidence, ``gold+silver'')}.
\end{equation}
To extract more detailed information, one needs to translate from $[\d
q/\d z]_0$ to jerk $j_0$.  Working to third order in redshift it is
straightforward to calculate ($k=0$):
\begin{equation}
d_L(z) = {c\;z\over H_0} \;
\left\{
1 + {1-q_0\over 2} z -{1-q_0^2+[\d q/\d z]_0\over 6} z^2 + O(z^3)
\right\},
\end{equation}
and so obtain
\begin{equation}
j_0 = q_0 + 2q_0^2 + \left.{\d q\over\d z}\right|_0.
\end{equation}
Using $w_0= -(1-2q_0)/3$ it is then easy to determine
\begin{equation}
\fl
w_0 \in (-1.2,-0.5); \qquad 
j_0 \in (-0.3,+5.9) \qquad
\hbox{(99\% confidence, ``gold'')},
\end{equation}
and
\begin{equation}
\fl
w_0 \in (-1.3,-0.5); \qquad 
j_0 \in (-0.1,+6.4) \qquad
\hbox{(99\% confidence, ``gold+silver'')}.
\end{equation}
Note that even at the 68\% level both data sets exhibit a thin sliver
of parameter space compatible with $q_0<-1$. (That is, $w_0<-1$, so
that the grand total of all stress-energy contributing to the
cosmological fluid is ``phantom''.)  Also note that the upper bounds
on the jerk are comparatively weak.

Now consider the linearized equation of state using
\begin{equation}
\kappa_0 = -{1\over3}\; {1-j_0\over 1+q_0}.
\end{equation}
This parameter is very poorly constrained
\begin{equation}
\kappa_0 \in (-\infty,-5.5) \cup (-0.5,+\infty) 
\qquad
\hbox{(99\% confidence, ``gold'')},
\end{equation}
and
\begin{equation}
\kappa_0 \in (-\infty,-4.5) \cup (-0.5,+\infty) 
\qquad
\hbox{(99\% confidence, ``gold+silver'')}.
\end{equation}
Note that very little of the real line is \emph{excluded}, and that
the entire positive axis is allowed by the data. Since this might at
first be a bit surprising, let me explain the basic reason for this
behaviour: If the permissible region contains a subset compatible with
$j_0\neq 1$ and $q_0=-1$, then this implies a subset of the
permissible region is compatible with $\kappa_0 = \pm\infty$. That is,
whenever there is a part of parameter space compatible with phantom
matter, then $\kappa_0 \to \pm\infty$ at the edge of the phantom
region.

Similarly, as long as the permissible region contains a subset
compatible with $j_0=1$ and $q_0\neq-1$, this implies a subset of the
permissible region is compatible with $\kappa_0 = 0$. Now $j_0= 1$
is equivalent to
\begin{equation}
\left.{\d q\over\d z}\right|_0 = (1+q_0)\;(1-2q_0)
\end{equation}
and it is easy to see that a subset of the Riess \emph{et al.}
permissible region is compatible with this constraint. With both of
these conditions being satisfied the entire positive real line is
allowed for $\kappa_0$. The portion of the negative real line attached
to $-\infty$ is associated with phantom matter, while the portion of
the negative real line attached to the origin is associated with
non-phantom matter ($w_0\geq -1$) with the peculiar property $\d
p/\d\rho < 0$.

It is also easy to see why at least some values of $\kappa_0$ are
excluded. If the permissible region contains the point $q_0=-1$,
$j_0=+1$ (equivalently $q_0=-1$, $[\d q/\d z]_0=0$) then, since this
corresponds to $\kappa_0$ taking on the indeterminate value $0/0$, it
is easy to convince oneself that for any open set surrounding this
point the computed values of $\kappa_0$ completely cover the entire
real line --- both positive and negative. But the point $q_0=-1$,
$j_0=+1$ ($q_0=-1$, $[\d q/\d z]_0=0$) is in fact excluded from both
gold and gold+silver datasets at more than 99\% confidence, which is
ultimately the reason that at least some values of $\kappa_0$ can be
excluded.  However, it should be emphasised that the constraints on
$\kappa_0$ are best described as extremely weak.

A slightly different analysis (using the same raw data analyzed in
somewhat different fashion) is presented by Caldwell and
Kamionkowski~\cite{Caldwell}.  Their result can usefully be summarized
(again assuming $k=0$) as:
\begin{equation}
q_0 \in (-1.1,-0.2); \qquad 
j_0 \in (-0.5,+3.9) \qquad
\hbox{(95\% confidence)},
\end{equation}
whence
\begin{equation}
w_0 \in (-1.1,-0.5)
\qquad
\hbox{(95\% confidence)},
\end{equation}
and
\begin{equation}
\kappa_0 \in (-\infty,-10) \cup (-0.6,+\infty) 
\qquad
\hbox{(95\% confidence)}.
\end{equation}
While the numbers are slightly different, the overall message is the
same: the jerk is relatively weakly constrained and $\kappa_0$ is very
poorly constrained.

If one wishes to constrain $\kappa_0$ to at least lie in some bounded
region of the real line, then one would need to seek improved data
that might exclude the possibility $q_0=-1$ ($w_0=-1)$. [This would
then be the death-knell for phantom matter.] Similarly, if one wishes
to constrain $\kappa_0$ to at least lie in the ``physically most
reasonable'' region (some bounded region of the \emph{positive} real
line), one would need to seek improved data that might exclude the
possibility $j_0< 1$. Obtaining a dataset of such quality would be
extremely challenging: Assuming no change in the location of the
center of the currently determined permissible region, this would
correspond to contracting the 99\% confidence intervals inwards to lie
somewhere inside the current location of the 68\% confidence
intervals.  This corresponds to approximately a three-fold decrease in
the standard deviations obtained when fitting the Hubble law to the
dataset; despite the advances in ``precision cosmology'', this appears
well beyond current (or even reasonably foreseeable) capabilities.

Furthermore, one should note that the above considerations are subject
to having additional data (or theoretical prejudices) available to
justify setting $k=0$ [or more precisely, $c/(H_0 a_0) \ll 1$].  In
the absence of such data, the third order term in the luminosity
distance Hubble law only implies~\cite{Caldwell}
\begin{equation}
\fl
q_0 \in (-1.1,-0.2); 
\qquad  
j_0 + {k\;c^2\over H_0^2 \;a_0^2} \in (-0.5,+3.9) \qquad
\hbox{(95\% confidence)}.
\end{equation}
Constraints on the equation of state, $\kappa_0$, are now even weaker.

Note that appealing to the quartic term in the luminosity distance
version of the Hubble law will not help, as that fourth order term
brings in an additional free parameter (the snap $s_0$), so that there
are still more free parameters than there are coefficients that can be
measured. To side-step this particular problem the two possibilities
are: (1) The traditional one, find additional data, above and beyond
the Hubble law, that somehow constrain the space curvature $k/a_0^2$.
(2) More challengingly, use a modified Hubble law that is independent
of space curvature $k/a_0^2$.  For instance, the ``physical distance
travelled'' version of the Hubble law $D(z)$ presented in equation
(\ref{E:physical}) is completely independent of space curvature ---
the challenge in this approach is to find a useful way of
observationally measuring $D$.

\section{Discussion}

There are currently many different models for the cosmological fluid
under active consideration. Though these models often make
dramatically differing predictions in the distant past (\eg, a
``bounce'') or future (\eg, a ``big rip'') there is considerable
degeneracy among the models in that many physically quite different
models are compatible with present day observations. To understand the
origin of this degeneracy I have chosen to rephrase the question in
terms of a phenomenological approach where cosmological observations
are used to construct an ``observed'' equation of state.  The key
result is that even at the linearized level, determining the slope of
the EOS requires information coming from the third order term in the
Hubble law. Unfortunately, while the experimental determinations of
the parameters appearing in the Hubble law are certainly improving, we
are still somewhat limited in what we can say concerning the
third-order term.  Despite the fact that some parameters in cosmology
are now known to high accuracy, other parameters can still only be
crudely bounded~\cite{precision}.  In particular, the jerk is
relatively poorly bounded, and as a consequence direct observational
constraints on the cosmological EOS (in the form of measurements of
$\kappa_0=[\d p/\d\rho]_0$) are currently extremely poor and are
likely to remain poor for the foreseeable future.

\appendix

\ack

This Research was supported by the Marsden Fund administered by the
Royal Society of New Zealand. I wish to thank T~Chiba, T~Padmanabhan,
A~Riess, and V~Sahni for their comments.

\section*{References}


\end{document}